\documentclass{an}

\usepackage{graphicx,fancyhdr}
\usepackage{times}

\begin{document}

\sloppy
\newcommand{\kms}{km\,s$^{-1}$}
\newcommand{\Halpha}{H$\alpha$}


\title{EURONEAR - Data Mining of Asteroids and Near Earth Asteroids}

      \author{O. Vaduvescu\inst{1,2,3}\fnmsep\thanks{\email{ovidiu.vaduvescu@gmail.com}\newline}, 
              L. Curelaru\inst{4}, 
              M. Birlan\inst{2,3}, 
              G. Bocsa\inst{3}, 
              L. Serbanescu\inst{3}, 
              A. Tudorica\inst{4,5}, 
              J. Berthier\inst{2}
              }

\titlerunning{EURONEAR - Data Mining of Asteroids and NEAs}
\authorrunning{O. Vaduvescu et al}

   \institute{Isaac Newton Group of Telescopes, 
              Apartado de Correos 321, E-38700 Santa Cruz de la Palma, Canary Islands, Spain \\
            \and
              IMCCE, Observatoire de Paris, 
              77 Avenue Denfert-Rochereau, 75014 Paris Cedex, France \\
            \and
              The Astronomical Institute of the Romanian Academy,  
              Cutitul de Argint 5, 040557, Bucharest, Romania \\
            \and
              The Romanian Society for Meteors and Astronomy (SARM), 
              CP 14 OP 1, 130170, Targoviste, Romania \\
            \and
              University of Bucharest, Department of Physics, 
              Platforma Magurele, Str. Fizicienilor nr. 1, CP Mg - 11, Bucharest Magurele 76900, Romania \\
             }

\date{Received; accepted; published online}



\keywords{asteroids, ephemerides, methods: data analysis, astronomical databases: miscellaneous}

\abstract{Besides new observations, mining old photographic plates and CCD image archives 
     represents an opportunity to recover and secure newly discovered asteroids, also to 
     improve the orbits of Near Earth Asteroids (NEAs), Potentially Hazardous Asteroids (PHAs) 
     and Virtual Impactors (VIs). These are the main research aims of the EURONEAR network.
     As stated by the IAU, the vast collection of image archives stored worldwide is still 
     insufficiently explored, and could be mined for known NEAs and other asteroids appearing 
     occasionally in their fields. This data mining could be eased using a server to search and 
     classify findings based on the asteroid class and the discovery date as ``precoveries'' 
     or ``recoveries''. We built PRECOVERY, a public facility which uses the Virtual Observatory 
     SkyBoT webservice of IMCCE to search for all known Solar System objects in a given 
     observation. To datamine an entire archive, PRECOVERY requires the observing log in a standard 
     format and outputs a database listing the sorted encounters of NEAs, PHAs, numbered and 
     un-numbered asteroids classified as precoveries or recoveries based on the daily updated 
     IAU MPC database. As a first application, we considered an archive including about 13,000 
     photographic plates exposed between 1930 and 2005 at the Astronomical Observatory in Bucharest, 
     Romania. First, we updated the database, homogenizing dates and pointings to a common format 
     using the JD dating system and J2000 epoch. All the asteroids observed in planned mode were 
     recovered, proving the accuracy of PRECOVERY. Despite the large field of the plates imaging 
     mostly $2.27\degr \times 2.27\degr$ fields, no NEA or PHA could be encountered occasionally 
     in the archive due to the small aperture of the 0.38m refractor insufficiently to detect 
     objects fainter than $V\sim15$. PRECOVERY can be applied to other archives, being intended as 
     a public facility offered to the community by the EURONEAR project. This is the first of a 
     series of papers aimed to improve orbits of PHAs and NEAs using precovered data derived from 
     archives of images to be data mined in collaboration with students and amateurs. In the next 
     paper we will search the CFHT Legacy Survey, while data mining of other archives is planned 
     for the near future. } 

\maketitle


\section{Introduction}
\label{intro}

A {\it Near Earth Asteroid} (NEA) is defined as an asteroid having a perihelion distance ($q$) 
less than 1.3 AU. A {\it Potentially Hazardous Asteroid} (PHA) is a NEA having a Minimum Orbital 
Intersection Distance (MOID) less than 0.05 AU and the absolute magnitudes ($H$) less than 22 mag, 
which corresponds to objects larger than 150m. This limit in size corresponds to asteroids large 
enough to potentially cause a global climate disaster and threaten the continuation of the 
human civilization (\cite{cha94}). A {\it Virtual Impactor} (VI) represents 
a NEA (mostly recently discovered) whose virtual orbit associated with the present observing 
uncertainty could collide with the Earth (\cite{boa03}). 

There are about 350 plate archives existing in the observatories around the globe holding more 
than two million photographic plates in their libraries. Created within the IAU Commission 9 by 
the Working Group on Wide Field Imaging, the Wide-Field Plate Database (WFPDB - www.skyarchive.org) 
contains the descriptive information for the astronomical wide-field ($>1\degr$) photographic 
observations stored in about 30\% of the estimated one thousand archives worldwide (\cite{tsv91,tsv05}). 
This wealth of astronomical information, specifically the WFPDB database, represents a great 
opportunity able to provide valuable scientific information for various astronomical objects 
observed occasionally in the past. 

In spite of this opportunity, as well as the exponentially increasing number of digital images 
acquired by large telescopes stored in archives at major observatories, very few projects have 
been devoted to data mining of asteroids. Probably the major obstacles for a wider use of the 
archives include: 1) the lack of easily pointings for different archives (observing logs available
in a homogeneous format, such as the WFPDB format accepted in VIZIER since 1997 
(http://webviz.u-strasbg.fr/viz-bin/VizieR?-source=VI/90); 2) the lack of the public tools 
(software and servers) to facilitate the data mining; and 3) the lack of digitized archives 
(much easier to access and search than traditional photographic plate archives). 

The Anglo-Australian Near-Earth Asteroid Survey (AANEAS) was one of the pioneering projects 
to use data mining as a tool to improve the orbits of NEAs (\cite{ste98}). AANEAS 
used both the UKST 1.2m Schmidt telescope to observe in survey mode, and its plate archive 
library to search for the newly objects which have appeared occasionally on the plates. 
Following the discovery of a new NEA and the determination of its orbit, a integration backward 
in time could be performed in order to compare its ephemerides against some existing unknown trails 
from the image archive. The possible identification of such an object is called a {\it precovery}, 
a term which is derived from ``pre-recovery'' - the act of recovering an object in the past 
(\cite{ste98}). 

During the last decade \cite{boa01} carried out in Italy the Arcetri NEO precovery 
program (ANEOPP). This project exploited mainly two archives in the search for new NEAs, 
namely the Arcetri plate library consisting in copies of the Palomar Observatory Sky Survey, 
the UKST southern sky survey, and the Digital Sky Survey. In only two years, the ANEOPP project 
was quite successful in precovering about 80 objects soon after their discovery, these becoming 
multi-opposition objects from single-opposition, with an orbit very much improved. A special 
focus was placed on NEAs in danger of being lost due to insufficient observations (e.g., the 
NEA 1998 NU), as well as on VIs in order to improve their orbits using precovered positions 
and remove their VI status (e.g. the PHA 2000 PN9). 

The DLR-Archenhold Near Earth Objects Precovery Survey (DANEOPS, \cite{hah02}) was 
initiated to systematically search existing photographic plate archives such as the Palomar 
Observatory Sky Survey (POSS - 1.2m Oschin Schmidt telescope) and the Anglo-Australian 
Observatory (1.2m UK-Schmidt telescope), i.e. using DSS 1 and 2 from the STScI Digitized 
Sky Survey. During about two years, this program precovered or recovered some 150 objects, 
mostly NEAs. 

On December 20, 2004 the Minor Planet Center announced the discovery of the famous 
asteroid 2004 MN4, namely Apophis (\cite{san08}). This object was 
discovered actually six months earlier by Tucker, Thollen and Bernardi from Kitt Peak, 
according to the data collected in two nights which presented some timing and astrometric 
problems. 
Just when the Christmas holidays were about to begin, it was already apparent that Apophis 
will become a VI PHA in 2029. According to the NEODyS/CLOMON2 system, this object was flagged 
to have an impact risk of TS=2 on the Torino Scale (\cite{bin00}) which was raised 
later to an unprecedented TS=4 and an infamous 1 in 38 chances of impact. This result was 
obtained based on more than 200 new observations gathered in 20 runs performed within one 
week following a call for new data which in fact aggravated the situation. Fortunately, 
the tense problem was solved using some precovery positions obtained by Gleason et al. 
(\cite{gle04}) based on some Spacewatch observations from March 2004 which extended the 
orbital arc to 9 months, ruling out any possible impact in 2029. Here is how only one precovered 
position could degrade a VI object to a PHA, ruling out a potential impact with the Earth, 
at least for a while. 

Created in 2006 in Paris, the European Near Earth Asteroids Research (EURONEAR) is a 
project which envisions to establish a coordinated network available to observe, discover 
and study NEAs and PHAs using 1-2m class telescopes, as well as data archives existent in 
photographic or digital means in various archives around the globe (\cite{vad08}). 
Via EURONEAR, we aim to increase the European synergies in a common network to provide 
specific tools and methods of research to the community willing to contribute in order 
to fight the asteroid impact hazard. In this first paper of this series, we present the 
software devoted to data mining of asteroids and its first application to the plate archive 
observed at the Astronomical Observatory in Bucharest, Romania. 
The paper is organized as follows: in Section~\ref{precovery} we present the method of 
search and PRECOVERY software. In Section~\ref{applications} we include four applications 
based on a few image archives. The first uses the plate archive acquired in Bucharest. 
In the Appendix we present the database associated to this archive, performing the data 
mining of known asteroids using PRECOVERY. 


\section{PRECOVERY Server}
\label{precovery}

Part of the EURONEAR project, we have envisioned to built a server aimed to data mine any 
astronomical archive for known NEAs and other asteroids. This facility will be available to 
the whole astronomical community willing to search image archives to which they have access. 
Within the future frame of the Virtual Observatory, basically all image archives could be 
linked in a common {\it Sky Mega-Archive} available for data mining using basic search 
capabilities such as PRECOVERY. Thanks to the easy access via Internet to the public archives 
and databases available nowadays, the interest in data mining for NEAs and other asteroids 
should be expected to grow soon. The work could involve not only professional astronomers, 
but also people working in the education, students, amateur astronomers and the public 
outreach. 

Built at IMCCE Observatoire de Paris, the SkyBoT tool was conceived to help astronomers 
in the data mining of the Solar System, to prepare and analyse astronomical observations, 
and to provide ephemerides of Solar System bodies in the frame of the Virtual Observatory 
(\cite{ber06}). The service uses a huge database (cca 2 TB) containing pre-computed 
ephemeris for all solar system objects, which is updated on a weekly basis. Using the capabilities 
of SkyBoT, an entire archive containing images (plates or CCDs) could be mined for precoveries 
and recoveries of known asteroids. The search could be implemented as a server which uses an 
input holding the observing log of the entire archive in a specific format. This should consist
in sets of observations (one line corresponding to one observation) to include the pointings 
given by the centres of the observed fields (RA, DEC) at epoch J2000, the instrument field of 
view (along the RA and DEC directions in degrees), the Julian Date JD (or calendar date and 
time UT), exposure time (in seconds), and the IAU observatory code. Another approach could
be to use the homogenous WFPDB format addopted by VIZIER. 

\begin{figure*}
\centering
\includegraphics[angle=0,width=14cm]{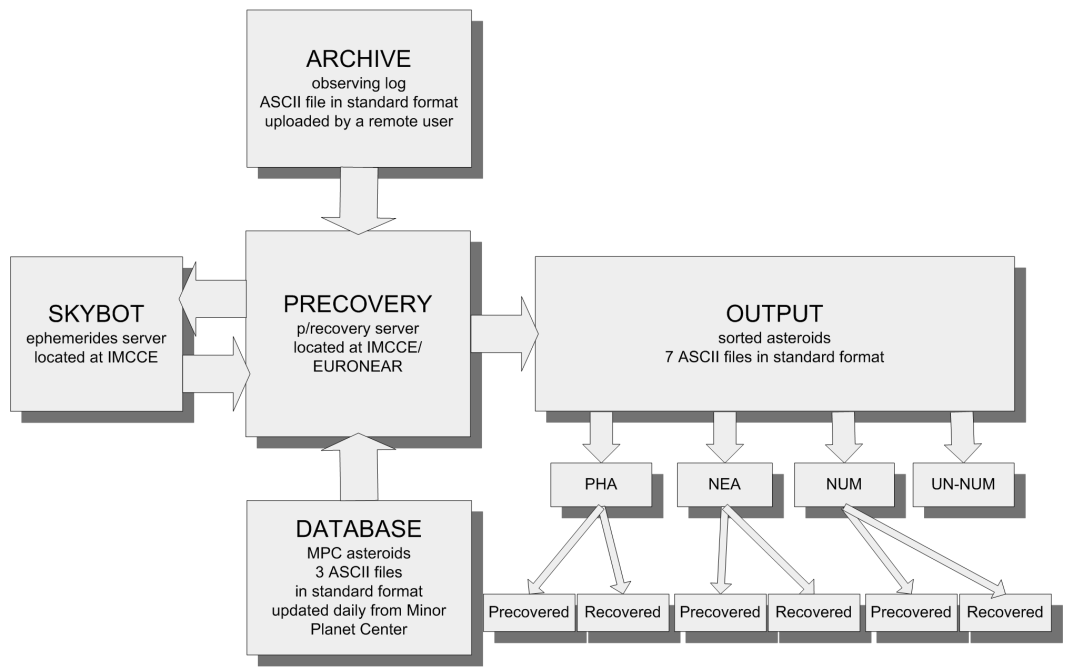}
\begin{center}
\caption{The flow chart of PRECOVERY data mining software, showing the method and the main interactions with other servers and databases. }
\label{fig1}
\end{center}
\end{figure*}

To perform the search, we wrote PRECOVERY, a PHP software which queries an entire archive of 
observations using SkyBoT, finds candidate images to hold asteroids appearing occasionally in 
the observed fields, and writes the output in a few lists to include PHAs, NEAs, numbered 
(NUM) and unnumbered (UNNUM) asteroids, sorting the results based on the asteroids' discovery 
date in two classes: {\it precoveries} (PREC - asteroids observed before their discovery date), 
and {\it recoveries} (RECOV - asteroids observed after their discovery). Some filters such as the 
limiting magnitude (set according to the telescope and instrument used) and filter could be 
applied based on the output of PRECOVERY in order to optimize the search based on specific 
science interests. The PRECOVERY software has been written in PHP in order to be easely embedded 
in the EURONEAR website (\cite{imc08}) and to be used in a network. 
In Figure~\ref{fig1} we include the flow chart of PRECOVERY, showing the dependencies with the 
SkyBoT server and MPC databases. 

To sort the findings given by SkyBoT, PRECOVERY uses the asteroids databases maintained by the 
Minor Planet Centre (MPC), namely the following lists of minor planets accessible from their 
{\it Unusual Minor Planets} website (\cite{iau08}): 1. List of Atens, Apollos, and 
Amors (from which we concatenate the list of known NEAs); 2. List of Potentially Hazardous 
Asteroids, and 3. List of Discovery Circumstances: Numbered Minor Planets. This database 
is updated by PRECOVERY to its server on a daily basis, in order to provide the most accurate 
information regarding the newest official denominations and orbits. 

The total running time of a search depends obviously on the length of the archive and the observed 
field of view. Currently, the server could not handle parallel querying of SkyBoT, this feature 
being possible to implement in a future version in order to decrease the execution time of large 
archives. In the current version of PRECOVERY, an average $2\degr \times 2\degr$ field takes about 
one minute for a complete search and selection. Given the repetitive queries to SkyBoT and the 
relatively long response time of this server which integrates orbits of hundreds or thousands of 
minor planets to distant epochs in time, it is expected to take up to a few hours or days for a 
complete search of an archive which includes hundreds or thousands of observations.
To optimize datamining of large archives, we recommend splitting the archive in more batches 
containing each up to 250 entries (observations). 

The output of PRECOVERY consists in 7 files to be downloaded by the user following the end of the 
run, which are stored for retrieval on the EURONEAR server temporarily. Besides the plate number, 
PRECOVERY output simply appends the output of SkyBoT, each line representing one occurrence (precovery 
or recovery event). Besides the asteroid number and name or its temporary designation, the table 
lists the asteroid type, ephemerides (RA and DEC at J2000), computed $V$ magnitude, the current 
ephemeris uncertainty for the requested epoch (CUE, in arcsec), and the apparent distance of the 
object from the plate centre (in arcsec). 


\section{Applications}
\label{applications}

\subsection{Bucharest Plate Archive}

The first archive available to us for this project started in 2005 was the photographic plate 
archive collected at the Astronomical Observatory in Bucharest since 1930 until present. This 
archive was acquired with the Prin Mertz double refractor (F=6m/D=0.38m) and holds about 13,000 
plates imaging mostly Solar System objects (asteroids, comets, planetary satellites), thus an 
area subtending mostly the ecliptic, imaged during the past eight decades (\cite{boc08}). In 
the direct focus of the astrograph, the 24cm $\times$ 24cm plates subtend 
$2.27\degr \times 2.27\degr$, thus a large field possible to hold asteroids. To data mine 
this archive with PRECOVERY, we had to update first the database to a common format. 
In the Appendix we present the archive and its associated databases, applying them to PRECOVERY. 

\subsection{Other Applications of Bucharest Plate Archive}

Besides searches of minor planets, the Bucharest plate archive could be used for other studies, 
such as to improve the orbits of double star systems imaged occasionally in the archive. Using 
the Bucharest 2007 plate database in a separate study, we searched the double star systems 
appearing on plates based on the Washington Double Stars catalog (\cite{cur07}). Many of these 
double systems have insufficiently observed orbits, so archive work is welcomed to improve these 
based on old observations (e.g., acc. to the Journal of Double Star Observations). A preliminary 
search using the Bucharest plate archive found about 2300 plates from the Bucharest archive to 
include measurable double stars in about 3600 systems. We plan to continue this work in a 
dedicated paper. 

\subsection{Canada-France-Hawaii Legacy Survey}

In a next paper we will apply PRECOVERY to search the Canada-France-Hawaii Legacy Survey (CFHTLS). 
This study has been conducted with positive results since 2007 (\cite{vad07}). 
Thanks to the large aperture of the 3.6m CFHT telescope equipped with the large $1\degr\times1\degr$ 
FOV of the MEGACAM camera, the PRECOVERY mining and image inspections have shown that about 250 NEAs 
and PHAs could be measured based on the CFHTLS archive, a work to be addressed in the second paper 
of this series. 

\subsection{Other Image Surveys}

Aimed as a public facility to be used within the EURONEAR network, PRECOVERY could be employed 
for data mining virtually any other archive in the world (plates or images), provided that the user 
has the access first to the observing log holding the archive in a standard format. Besides the 
Bucharest plate archive and the CFHTLS image archive, we intend to apply PRECOVERY to mine other 
image archives acquired on large telescopes, in a distributed team effort to include also students 
and amateurs to work within the EURONEAR network. In a first instance, one requires only the 
observing logs corresponding to the archive in order to run the PRECOVERY search, then the access 
to only a few selected candidate images expected to hold the objects of interest. 

At least four other archives have been identified for data mining of the most important NEAs, 
namely those in great need of data such as PHAs, VIs, and lost NEAs. First, we plan to extend our 
initial CFHTLS study to comprise the entire Megacam archive (FOV = $1\degr \times 1\degr$) acquired 
with the CFHT 3.6m telescope and maintained online by the Canadian Astronomical Data Center (CADC). 
Second, we aim to mine the Wide Field Survey taken with the Wide Field Camera (WFC, 
$34\arcmin \times 34\arcmin$) of the 2.5m Isaac Newton Telescope in La Palma, Canary. 
Third, we plan to use the Wide Field Imager archive ($34\arcmin \times 33\arcmin$) acquired the 
ESO/MPG 2.2m telescope in la Silla, Chile and maintained online by ESO. Fourth, we plan to use the 
OGLE-III archive ($35\arcmin \times 35\arcmin$) acquired with the Warshaw 1.3m telescope in Las 
Campanas, Chile. All these instruments have a small pixel size (around 0.3 arcsec/pixel) and a 
very large field of view, being very adequate to ensure accurate astrometry for many NEAs in 
great need of data appearing occasionally in the field. 


\section{Conclusions}
\label{conclusions}

Besides new observations, mining the old photographic plates and CCD image archives represents 
one of the most important opportunity of research for improving the orbits of NEAs, and one of 
the aims of the EURONEAR project. 

There are quite many old wide field plate archives around the globe, as counted by the WFPDB. 
This huge collection of astronomical information is still insufficiently explored, being possible 
to be data mined quite easily for known NEAs, PHAs, VIs, and other asteroids possible to appear 
occasionally in the observed fields. The resulted findings could be sorted as {\it precoveries} 
(observed before the discovery of a given object) or {\it recoveries} (observed after discovery). 
The resulting astrometry could bring important contributions in the direction of the NEA research 
and the PHA impact hazard. 

In this sense, we created PRECOVERY as a public facility integrated within the EURONEAR project. 
PRECOVERY is a service which exploits the SkyBoT facility of IMCCE to search for all known Solar 
System objects for a given observation. Embedded within the EURONEAR website, PRECOVERY requires 
an observing log of a given archive to search, and the IAU observatory code. The output consists 
of seven ASCII files which list the occasional encounters of NEAs, PHAs, numbered and un-numbered 
asteroids, classifying them as precovery or recovery based on the daily updated IAU minor planet 
MPC databases. 

In this paper we applied PRECOVERY to the plate archive taken at Bucharest Observatory using 
the 0.38m Prin Mertz astrograph during the last eight decades, which used mostly 24cm $\times$ 24cm 
plates imaging a relatively large $2.3\degr \times 2.3\degr$ field of view. Because of the small 
telescope aperture, insufficiently to image objects fainter than $V\sim15$, no PHA or NEA appearing 
occasionally on the plates were found in the archive. Nevertheless, all numbered asteroids and 
about 100 NEAs observed in the planned mode were precovered or recovered, proving the accuracy of 
PRECOVERY. 

Other applications of PRECOVERY and the Bucharest plate archive could be used for similar searches. 
In a preliminary search using the same Bucharest plate database, we found about 2300 plates from the 
Bucharest archive to include about 3600 double star systems, possible to be used to improve the 
orbits of this systems. Another search using PRECOVERY applied to Canada-France-Hawaii Legacy Survey 
resulted in about 250 images to hold NEAs and PHAs measurable. Both these, as well as mining other 
archives for NEAs, will be address by our future work. 


\begin{acknowledgements}
We acknowledge the Astronomical Institute of the Romanian Academy for providing access to the 
plate database, also for the scanning necessary to perform the limiting magnitude test. The study 
made use of SkyBoT, a web service developed by IMCCE Observatoire de Paris. Milcho Tsvetkov 
(The Institute of Astronomy in Sofia, Bulgaria, and the Working Group on Wide Field Imaging of 
the Commission 9 of the IAU) provided us some feedback and encouragements in this project. 
To measure the photometry and query the USNO-A2 catalog we used the VizieR and Aladin services 
at Centre de Donnees astronomiques de Strasbourg. 
\end{acknowledgements}

\appendix

\section{Bucharest Plate Archive}

\subsection{Overview of the Archive}

Celebrating its centenary in 2008, the Astronomical Observatory in Bucharest is affiliated 
today to the Astronomical Institute of the Romanian Academy (\cite{air08}). Despite the 
growing sky pollution due to its location in the capital Bucharest (2.5 million people), the 
observatory has contributed with many astrometric catalogs of stars, photographic observations 
of asteroids, comets, satellites, planets and stars observed during eight decades. The 
observatory has the IAU code 073. 

One of the main telescopes of the observatory has been the Prin Mertz double refractor 
(F=6m/D=0.38m; WFPDB code BUC038). This equatorial astrograph was endowed with a small 
piggy-back Zeiss Canon photographic camera (F=0.8m/D=0.16m; WFPDB code BUC016) used for 
wider field imaging during the first two decades. 
From 1930 until 1951, photographic plates 13cm $\times$ 18cm were used in the focus of both 
refractors, providing a field of view of $1.70\degr \times 1.23\degr$ for the Prin Mertz 
refractor and $12.00\degr \times 8.67\degr$ for the Zeiss Canon camera. In 1952, a new 
numbering system was put in place to accommodate the observations using larger 24cm $\times$ 24cm 
plates which provided a field of view of $2.27\degr \times 2.27\degr$ in the focus of the Prin 
Mertz refractor. In total, during the last 78 years, about 13,000 plates have been acquired in 
all configurations using both refractors, from which about 12,800 plates have been taken using 
the Prin Mertz astrograph and only 180 plates using the Zeiss Canon camera. 

In the past, the astrometric positions of the observed objects were measured manually with 
an ASCORECORD machine in the frame of stars from SAO or PPM catalogs. Recently, about half of 
the entire plate archive were digitized at low resolution to be available for preview part of 
the WFPDB collaboration with the Working Group on Wide Field Imaging of the IAU (\cite{boc08}). 
A few plates were scanned and digitized recently at high resolution (2400 dpi) 
using a scanner UMAX Alpha Vista II, although no plan to scan the entire archive exists yet. 

The Bucharest plate archive includes three related databases holding the observing logs of 
all observations performed at the observatory since 1930. The first database is updated monthly 
by the observatory in an ASCII format (\cite{boc08}), while the second included observations 
between 1930 and 1995 in a DBASE format (\cite{vas94a}). In 2007 we transformed the original ASCII 
database in a new database in MS Excel, which homogenizes the date and time in the standard JD 
format, transforming positions to a common J2000 epoch. We present next the three databases 
associated with the Bucharest archive. 

\subsection{The Original Database (\cite{boc08})}

Every last photographic plate from the library was checked and counted in 1994, then entered in 
an ASCII database which includes the observing logs holding the following columns: the plate number, 
pointing of the telescope (RA, DEC), epoch (in a non-standard format, namely, B1900, B1930, B1950, 
J2000, etc), plate size (in cm), observing date and time (in a non-standard format, namely local 
sidereal time), number of exposures, exposure time, and the observed object. It would be difficult 
to work with this database, given its non-standard format to express the date/time and pointing of 
the fields, so we had to transform it first. 

\subsection{The 1994 Database (\cite{vas94a})}

\cite{vas94a} describes the wide-field plate archive database taken in Bucharest 
since 1930 until 1994, while \cite{vas94b} makes a preliminary analysis of this database. 
Most observations focused on astrometry of minor planets ($78\%$), followed by fundamental stars 
($9\%$), comets and nebulae ($4\%$ each), and other objects (variable stars, radio sources, star 
clusters, double stars, major planets, etc). The exposure time were mostly less than 20 min (with 
less than $10\%$ plates exposed as long as about one hour), thus the limiting magnitude is expected 
to be quite reduced given the small 0.38m refractor observing from a big city, namely $V\le15$ at
its best. In fact, mostly due to the growing light pollution of Bucharest, the observations 
decreased during the last two decades, so the above figures regarding the surveyed objects did 
not change much. 

\subsection{The 2007 Database (the present work)}

\begin{figure}
\centering
\includegraphics[angle=0,width=8cm]{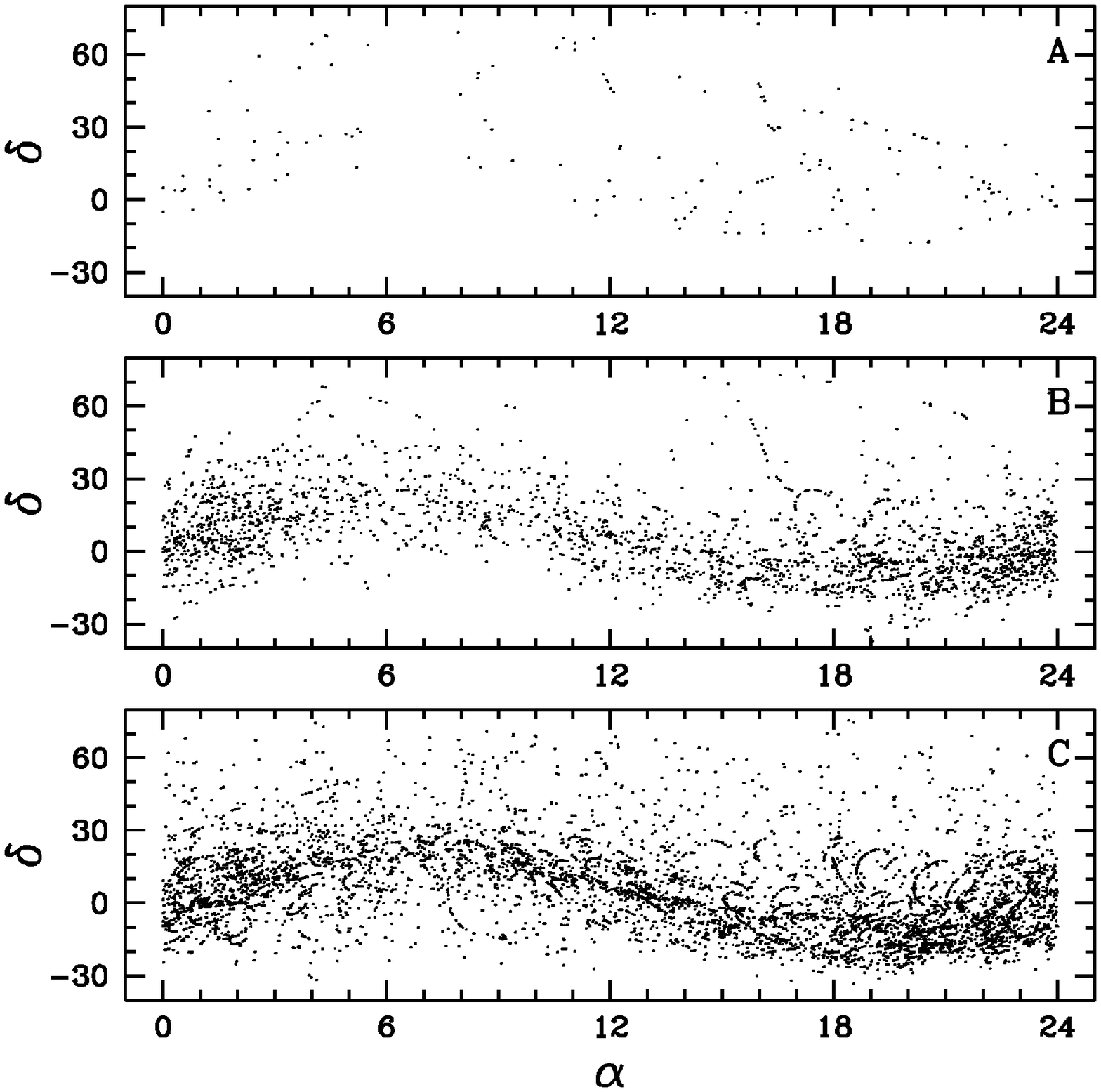}
\begin{center}
\caption{Distribution of the centres of the Bucharest plate archive on the sky. 
Panel A: $12.00\degr \times 8.67\degr$ plates observed with the Zeiss Canon photographic camera (179 plates); 
Panel B:  $1.70\degr \times 1.23\degr$ plates observed with the Prin Mertz refractor (2708 plates); 
Panel C:  $2.27\degr \times 2.27\degr$ plates observed with the Prin Mertz refractor (7824 plates). 
The region of the ecliptic could be evidenced in all plots, being traced by most asteroids and comets observed 
between 1930 and 2005. 
}
\label{fig2}
\end{center}
\end{figure}

Likewise the original ASCII database, the 1994 database created by G. Vass in DBASE conserves the 
original date and time recorded in local sidereal time, which is inconvenient for long time studies 
of the archive. Given this, we imported the original ASCII 2007 database in an MS Excel database, in 
order to facilitate a convenient access to the archive. First, we converted the observing dates and 
time to Julian Date (JD). Second, we transformed the original positions holding the telescope pointings 
expressed at different epochs to the common J2000 frame, correcting these by precession and nutation 
(\cite{mee99}). Third, we included the field of view (FOV), taking into account the plate 
scale based on the instrument configuration (two refractors and two type of plates). Finally, we 
added the exposure times in case of multiple exposures of the same plate. Most of this work was 
done automatically in MS Excel in Windows and Gnumeric in Linux, doubled finally by some inspection 
and manual editing. A few entries (about 100) failing to specify the observing time were deleted. 
The final Gnumeric database was exported to two ASCII files which list 1906 observations before 
1952 (the old series), and 8,805 observations after 1952 (the new series), adding together 10,711 
observations. 

In Figure~\ref{fig2} we plot the sky distribution of the Bucharest plate archive, according to the 
2007 database. In the top panel A we include centres of the exposures using the 13cm $\times$ 18cm 
plates exposed in the focus of the Zeiss Canon photographic camera (179 plates). In panel B we 
include the exposures of the 13cm $\times$ 18cm plates in the focus of the Prin Mertz astrograph 
(2,708 plates). Finally, in panel C we include the exposures of the 24cm $\times$ 24cm plates in the 
focus of the Prin Mertz astrograph (7,824 plates) which constitutes $73\%$ of the entire database. 

In the electronic Table~\ref{tbl-1} and Table~\ref{tbl-2} we include the 2007 database. The 
first table lists all observations before 1952 (the old series), and the second table includes all 
observations after 1952 (the new series). The new format includes the plate number, a flag listing 
the series, Julian Date (at start of exposure), the exposure time (in seconds), the plate centers 
at J2000 (RA in hours with decimals and DEC in degrees with decimals), the field of view (in the 
direction of RA and DEC, both in degrees), and the observed object (alpha-numeric). All fields are 
separated by the ``$|$'' character to ease the import of the fields.

\subsection{Limiting Magnitude}
\label{lim_mag}

Due to the small aperture of the Prin Mertz refractor (D=0.38m), also to the growing sky pollution
in Bucharest, the limiting magnitude is reduced to about $V<15$ (\cite{boc08}). Actually, this 
limit has been reached only for a few hundred plates observed in the best conditions: the longest 
exposures, good weather conditions, low airmass, less light pollution, good developing process, and 
good archival conditions. For most of the plates in the archive, the limiting magnitude is estimated 
to only about $V<12$ (\cite{boc08}). 

To evaluate the limiting magnitude of the plates observed in the the best conditions, we inspected 
8 plates scanned at 2400 dpi. The original TIFF images were transformed to FITS format for a smaller 
field ($46.2\arcmin \times 46.2\arcmin$) located at the centre of the plate, selected to maximize 
the image quality which is better near the centre. From the 8 plates, only one observed in 1965 
(nr. 5089) was deep enough (exposure time 30 min) to produce a good quality FITS image. We inspected 
this image in IRAF using the DAOPHOT package (DAOFIND task to find all stars within $5\sigma$ detection 
limit, and PHOT to measure their magnitudes). We measured the instrumental magnitudes using 20 selected 
stars from the centre of the field whose apparent $B$ magnitudes were taken from the USNO-A2 catalog 
(\cite{mon98}). Their measured errors were around $\pm 0.2$ mag, consistent with the catalog 
photometric errors. 

\begin{figure}
\centering
\includegraphics[angle=0,width=8cm]{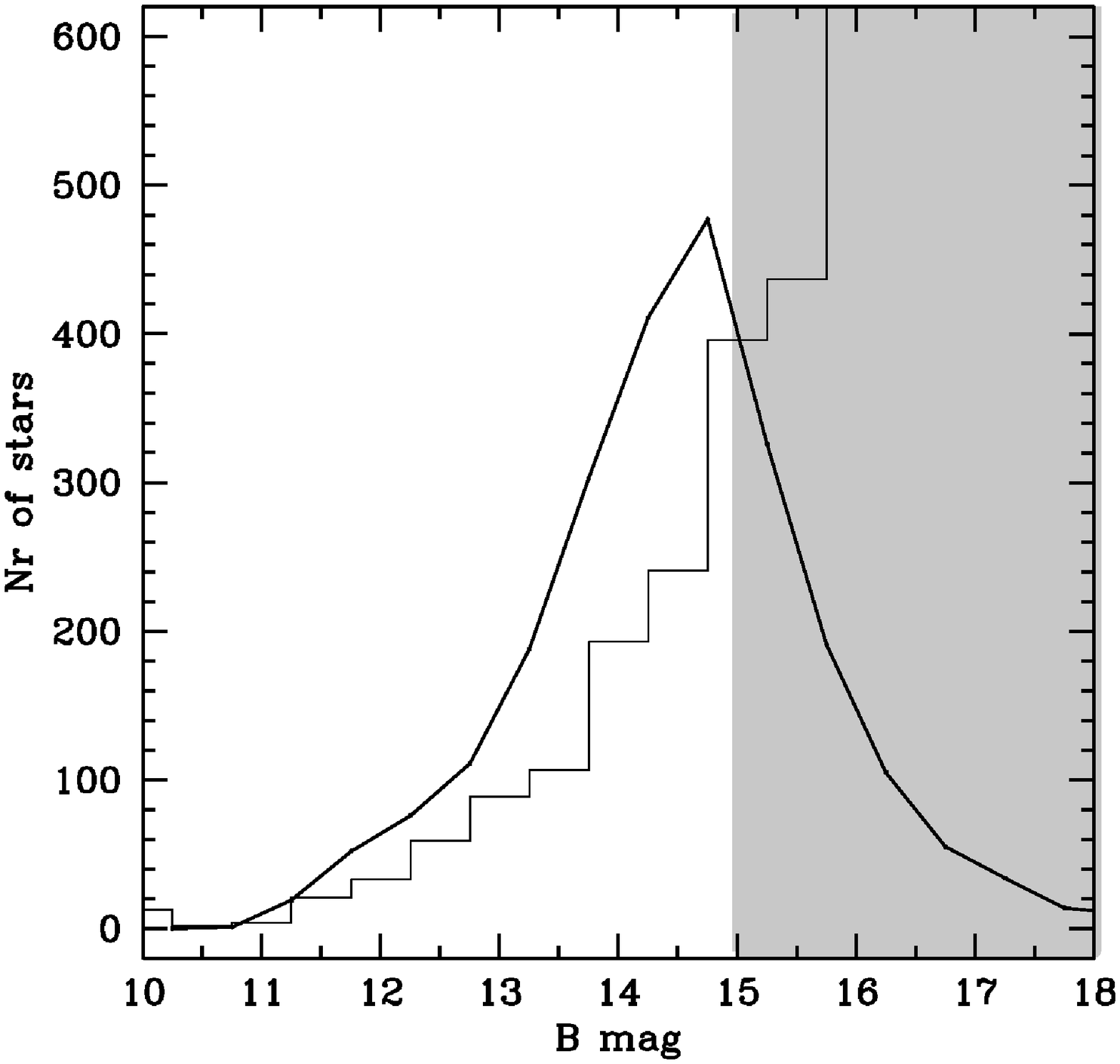}
\begin{center}
\caption{The distribution of the observed $B$ magnitudes of stars measured in the field of the plate nr. 5089 
(exposure time 30 min). The peak of the histogram combined with the detection of stars having known USNO-A2 $B$ 
magnitudes suggests a limiting magnitude $B\sim15.5$. This corresponds to $V\sim15$ at most, assuming $0.5<B-V<1$ 
for the entire asteroid population, reached only for plates observed, processed and stored in the best conditions. } 
\label{fig3}
\end{center}
\end{figure}

\begin{figure}
\centering
\includegraphics[angle=0,width=8cm]{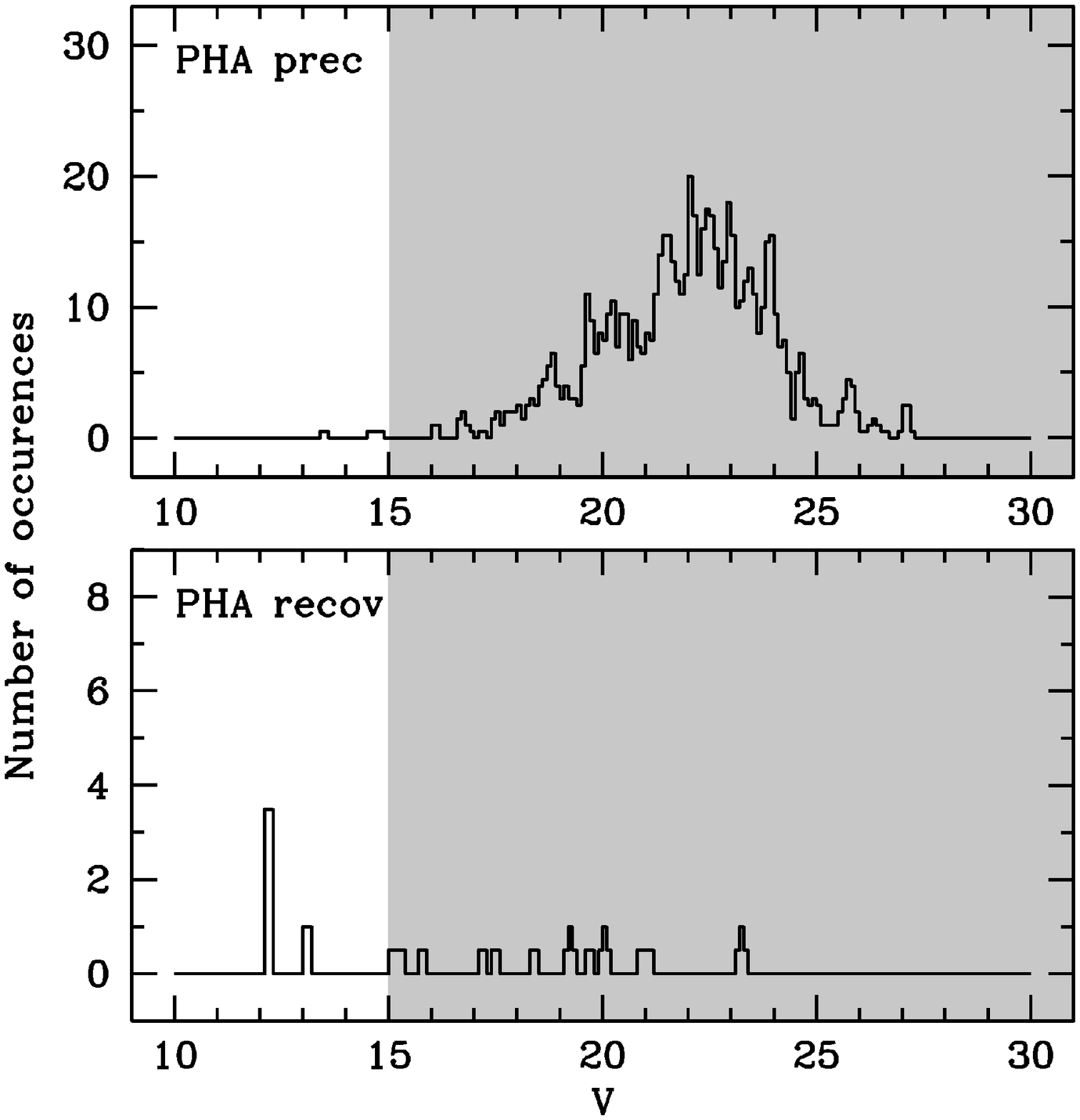}
\begin{center}
\caption{The distribution of the findings of PHAs from the Bucharest plate archive. 
The region in gray was inaccessible to the refractor. }
\label{fig4}
\end{center}
\end{figure}

\begin{figure}
\centering
\includegraphics[angle=0,width=8cm]{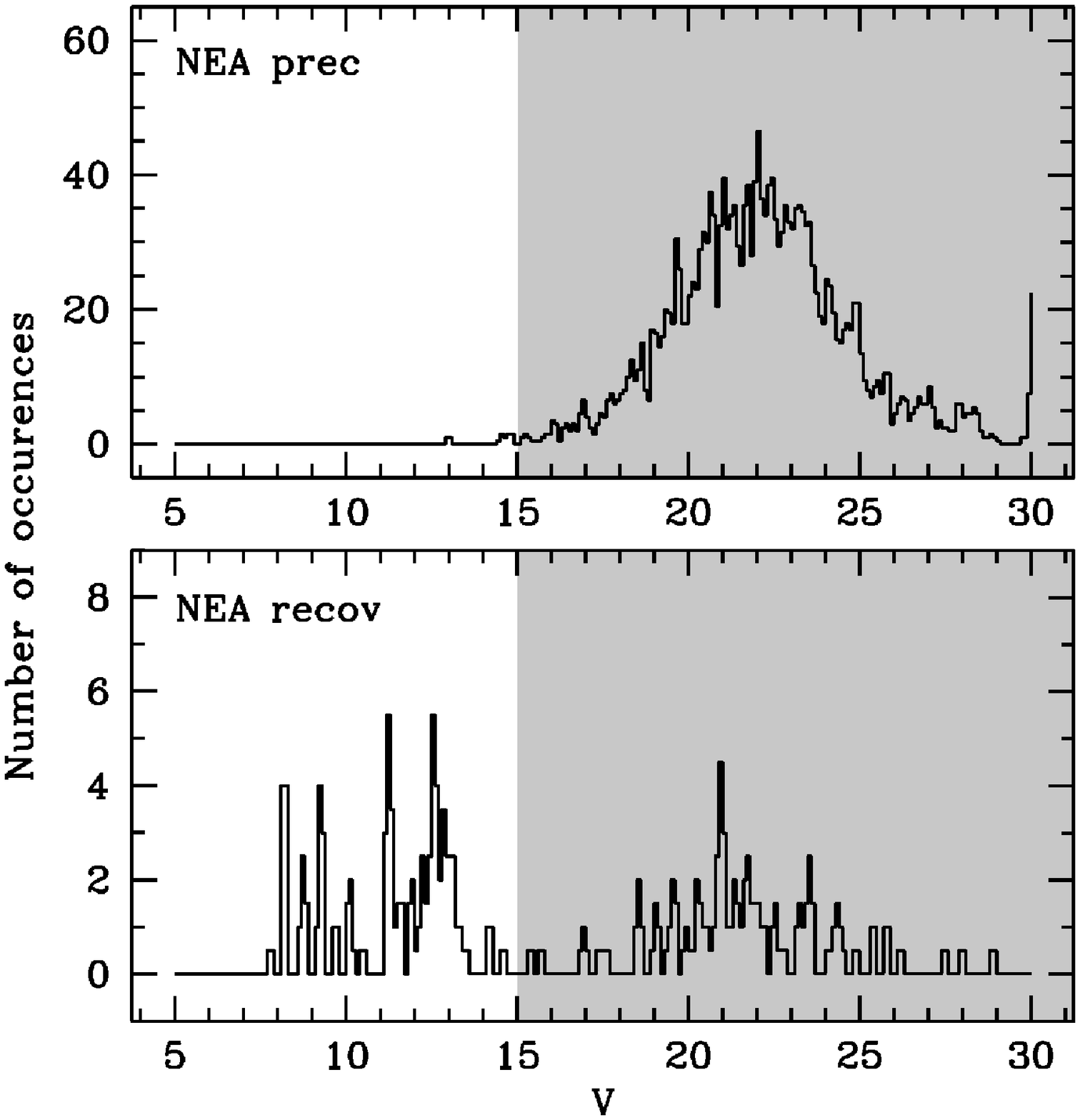}
\begin{center}
\caption{The distribution of the findings of NEAs from the Bucharest plate archive. 
The region in gray was inaccessible to the refractor. }
\label{fig5}
\end{center}
\end{figure}

In Figure~\ref{fig3} we present the histogram (plotted as a solid line) showing the number 
of sources detected by DAOPHOT in the scanned image as a function of the apparent $B$ magnitude. 
The shallow decrease of the curve could be related to about two reasons. The first relates with 
the detection of artifacts instead of real stars towards the faint limit, appearing most probably 
from the developing, archiving, and scanning processes. The second possible reason could be related 
with the comparison of magnitudes: while USNO-A2 catalog reports $B$ magnitudes calculated from the 
original photographic magnitudes, no filter was used for the observations in Bucharest, therefore 
the brightness of sources detected on plates could be over evaluated. To check this effect further, 
we considered the USNO-A2 stars within the same exact image, whose scaled histogram was compared 
with the linear one of the plate in Figure~\ref{fig3}. There is a small apparent shift between 
the two histograms (around 0.5 mag) which appears to grow following $B\sim13$, suggesting the two 
possible reasons. The plate histogram peaks around $B\sim15$, although the limiting magnitude 
could reach $B\sim15.5$, because we could identify stars as faint as this limit. Assuming an 
average colour of $0.5 < B-V < 1.0$ for most asteroids, it is prudent to consider the limiting 
magnitude $V=15$ for the entire plate archives, but keep in mind that this is possible to reach 
only in the best conditions. 

\subsection{PRECOVERY Mining}

We performed the data mining of the Bucharest plate archive for the new series database (1952-2005). 
We searched this archive only because SkyBoT currently does not accept epochs earlier than 1 Jan 1950. 
A future version of SkyBoT to be able to search epochs before 1950 requiring an update of the 
hardware (additional 2 TB) is expected to be implemented soon. A total of 8,805 plates observed 
between 1 Jan 1952 and 30 Sep 2005 with FOVs mostly of $2.27\degr \times 2.27\degr$ acquired with the 
0.38m Prin Mertz refractor were considered for the search. The complete search took about one month 
and about 50 batches executed in separate runs (each holding about 200 plates on average). To perform 
the search, we used the MPC asteroid database published on 6 Feb 2008. Given the shallow limiting 
magnitude reached by the archive, most asteroids discovered after this date will be too faint to 
appear on the plates, so one could consider this search almost complete. 

Table~\ref{tbl-3} shows a sample of the final list which includes the findings of the precovered 
PHAs searched in the 2007 Bucharest plate archive, listed without any filter regarding the limiting 
magnitude. The complete databases includes 7 ASCII files and is available in electronic form upon request. 
Table~\ref{tbl-4} lists the possible number of findings as a function of the asteroids' magnitude. 
In the first column we give the total number of findings found by PRECOVERY based solely on the 
asteroids' predicted positions by SkyBoT, with no filter of magnitude. Many PHAs, NEAs, NUMs and 
UNNUMs are located within the field of the plates in the archive, given the large field of view used. 

\begin{figure}
\centering
\includegraphics[angle=0,width=8cm]{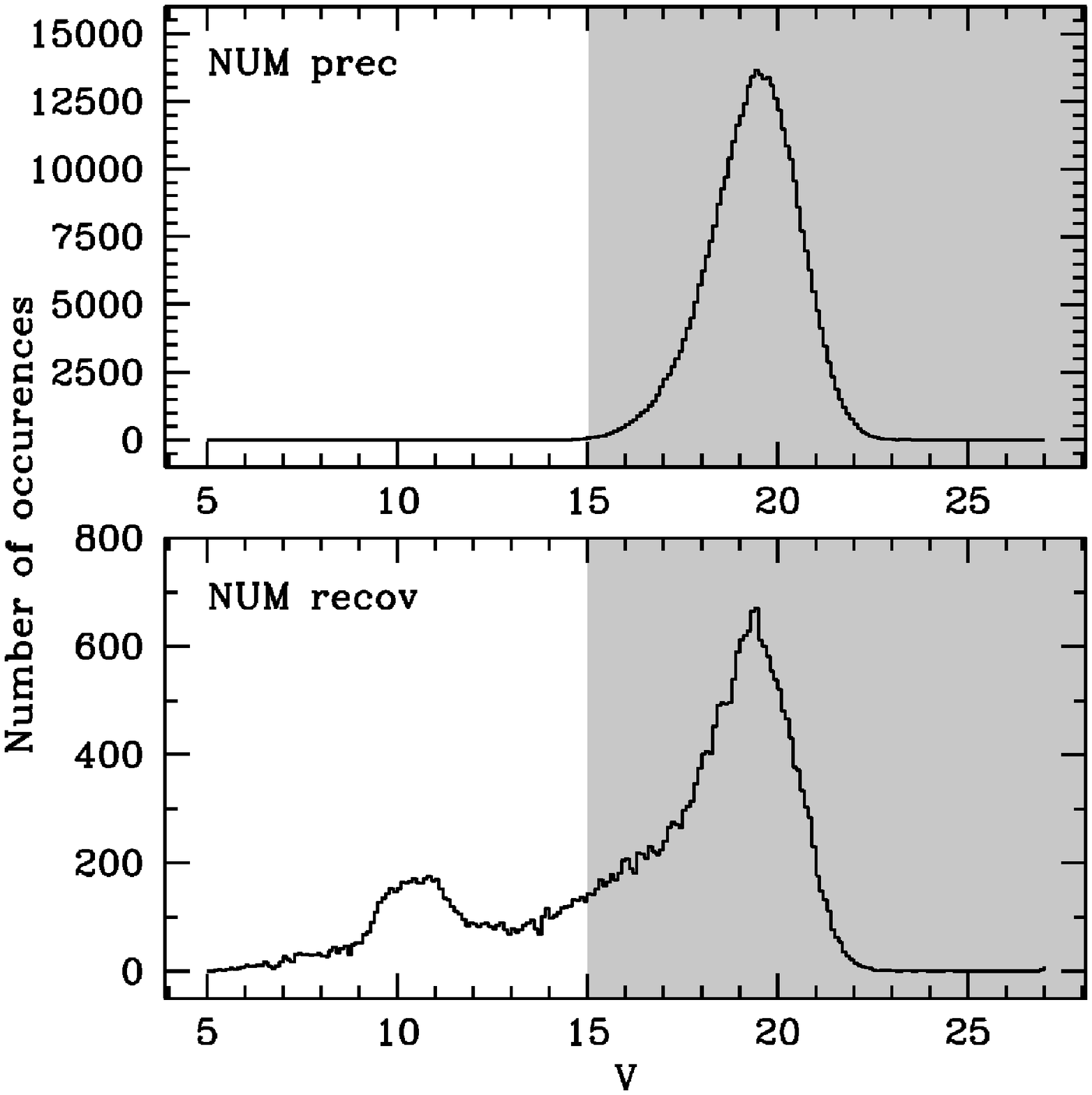}
\begin{center}
\caption{The distribution of the findings of numbered asteroids from the Bucharest plate archive. 
The region in gray was inaccessible to the refractor. }
\label{fig6}
\end{center}
\end{figure}

\begin{figure}
\centering
\includegraphics[angle=0,width=8cm]{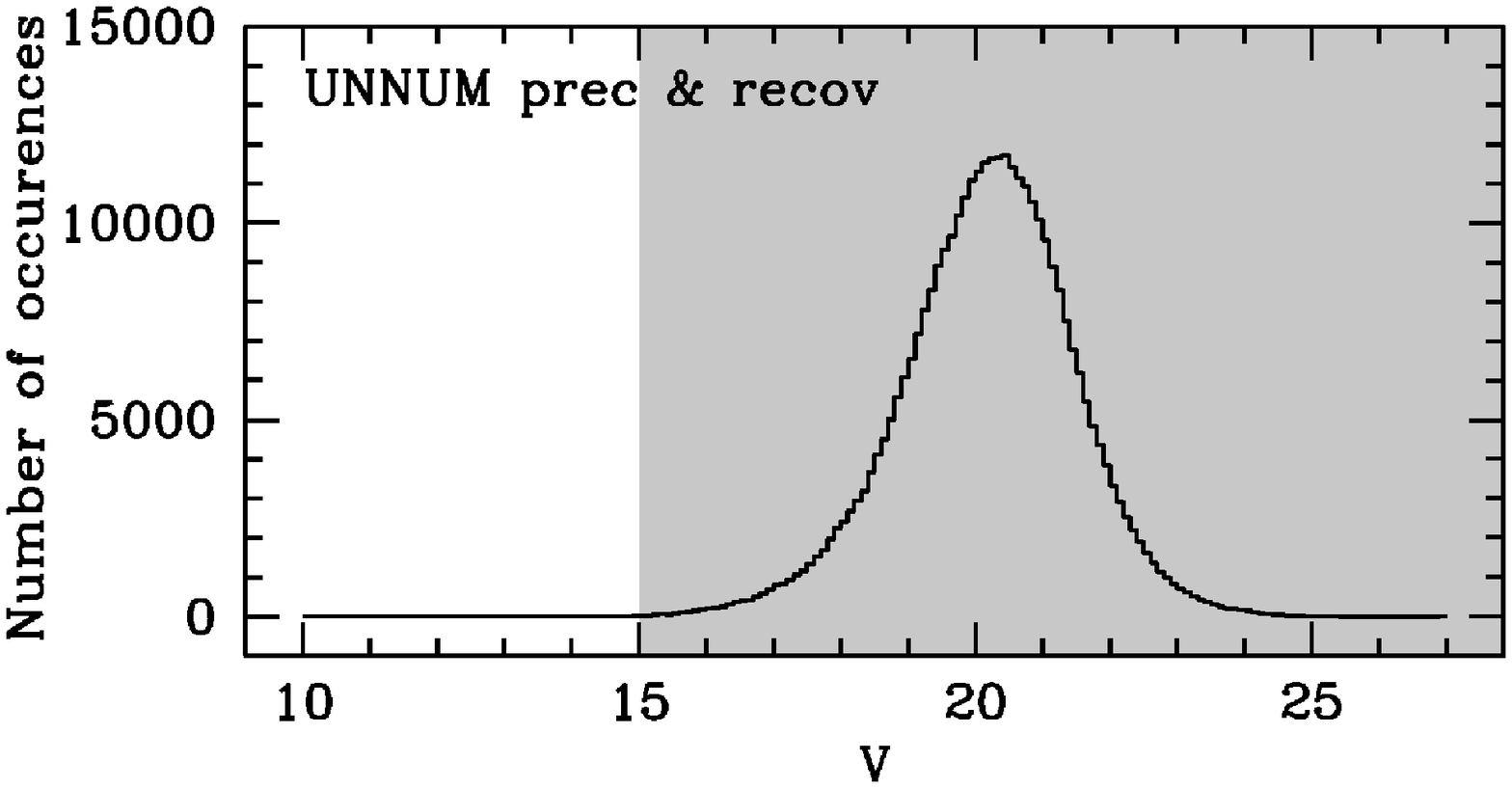}
\begin{center}
\caption{The distribution of the findings of unnumbered asteroids from the Bucharest plate archive. 
The region in gray was inaccessible to the refractor. }
\label{fig7}
\end{center}
\end{figure}

In Figures~\ref{fig4}, \ref{fig5}, \ref{fig6} and \ref{fig7} we show the distribution of the PHAs, 
NEAs, numbered (NUM), and unnumbered (UNNUM) asteroids encountered by PRECOVERY to be located in some 
candidate fields, independently of the magnitude limit. The gray regions refer to invisible objects, 
according to the considered limiting magnitude $V=15$. Although the astrograph was too small to image 
objects fainter than $V\sim15$, all distributions show the potential of similar precovery searches 
using a similar FOV with larger telescope aperture. 

As one could expect, the overwhelming majority of the findings list faint or very faint objects, 
inaccessible to the small 0.38m refractor used in Bucharest. To select the candidate plates holding 
findings possible visible in the archive, we ordered the database taking into account the asteroids' 
predicted magnitude by SkyBoT. The second column of Table~\ref{tbl-4} lists the number of candidate 
plates holding possible findings of asteroids with magnitudes $V<15$. 105 encounters of NEAs and PHAs 
were found (both precoveries and recoveries). Checking these findings versus the original database, 
one could observe that the majority refer to objects observed in the planned mode. These objects 
were recovered by PRECOVERY in all cases, probing the validity of our search. 

Considering $V=15$ as the overall limiting magnitude of the archive (see Section~\ref{lim_mag}), 
we expect up to 3 PHAs possible to precover and 2 PHAs possible to recover observed in planned mode 
in 9 findings. Moreover, 6 NEAs are expected to be precovered in 8 findings and 7 NEAs to be recovered 
in 85 findings (most observed in planned mode, and only 3 objects occasionally). We include in 
Table~\ref{tbl-5} all possible findings (PHAs and NEAs up to the limiting magnitude $V=15$). Following 
the inspection and scanning of these plates, we concluded that all the remaining 12 findings (NEAs and 
PHAs precovered or recovered occasionally) are actually invisible.

Due to the shallow magnitude limit of the refractor, very few minor planets are actually visible in 
the archive, most being numbered asteroids. As one can see in Figures~\ref{fig4} and \ref{fig5}, there 
are more encounters of precovered PHAs and NEAs than recoveries, which is normal given that the majority 
population of NEAs/PHAs has been discovered during the last two decades, compared with the archive 
which was mostly productive before the last period. One can observe two bulks in the NEAs recovered 
distribution, the first around $V\sim11$ corresponding observed objects in the planned mode, and the 
other around $V\sim22$, which coincides with the only peak in the NEA and PHA precovered distribution, 
corresponding to the limit of the present NEA surveys.

As one can see in Figure~\ref{fig6}, the peak of the precoveries and recoveries of numbered asteroids 
is located around $V\sim19.5$, while the peak of occasional observations of unnumbered asteroids is 
around $V\sim20.5$ as can be observed in Figure~\ref{fig7}, both being accessible by 1m class telescopes. 
Two main bulges can be observed in Figure~\ref{fig6}, corresponding to recoveries of numbered 
asteroids, with the first between $5 \le V \le 13$ corresponding to minor planets observed in planned 
mode. A clear cut in the distribution of the magnitudes exists around $V=13$, suggesting that this was 
the actual limiting magnitude possible using the given equipment.

\begin{table*}[!t]
\begin{center}
\caption{Sample of the Bucharest 2005 plate archive database including the old series 
         of observations (1930-1951). The complete database is available in electronic form upon request. } 
\label{tbl-1}
\begin{tabular}{lrrrrrrl}
\hline
\hline
\noalign{\smallskip}
Plate nr. & JD & Exp time & $\alpha$ centre & $\delta$ centre & FOV & FOV & Object \\
(old series) & (start exposure)  & (sec) & (J2000) & (J2000) & $\alpha$ $(\degr)$ & $\delta$ $(\degr)$ & \\
\hline
\noalign{\smallskip}
1 & 2426113.44442028 & 2400 & 17.20714 & 37.0510 & 12.00 & 8.67 & COMET 1930D \\
16 & 2426304.44093162 & 4200 & 5.28980 & 28.0757 & 1.70 & 1.23 & (198) AMPELLA \\
44 & 2426449.41968583 & 900 & 13.98881 & -7.7181 & 1.70 & 1.23 & (236) HONORIA \\
75 & 2426488.36860231 & 720 & 16.99918 & -15.9527 & 1.70 & 1.23 & (8) FLORA \\
83 & 2426511.39929546 & 3660 & 18.36830 & -9.9328 & 12.00 & 8.67 & (107) CAMILLA \\
\hline
\hline
\end{tabular}
\end{center}
\end{table*}

\begin{table*}[!t]
\begin{center}
\caption{Sample of the Bucharest 2005 plate archive database including the new series 
         of observations (1952-2005). The complete database is available in electronic form upon request. } 
\label{tbl-2}
\begin{tabular}{lrrrrrrl}
\hline
\hline
\noalign{\smallskip}
Plate nr. & JD & Exp time & $\alpha$ centre & $\delta$ centre & FOV & FOV & Object \\
(new series) & (start exposure)  & (sec) & (J2000) & (J2000) & $\alpha$ $(\degr)$ & $\delta$ $(\degr)$ & \\
\hline
\noalign{\smallskip}
1 & 2434027.23065004 & 720 & 6.31462 & 0.9453 & 1.70 & 1.23 & (41) DAPHNE \\
9 & 2434042.24232689 & 1080 & 7.73720 & 20.2960 & 1.70 & 1.23 & (407) ARACHNE \\
1135 & 2435723.26344971 & 2340 & 18.50261 & -10.1485 & 2.27 & 2.27 & (3) JUNO \\
1139 & 2435723.48645027 & 2520 & 23.41830 & -11.9084 & 2.27 & 2.27 & (40) HARMONIA \\
1288 & 2435858.27297292 & 3000 & 3.33999 & 16.0132 & 2.27 & 2.27 & (42) ISIS \\
\hline
\hline
\end{tabular}
\end{center}
\end{table*}

\begin{table*}[!t]
\begin{center}
\caption{Sample of findings representing precovered PHAs in the Bucharest 2005 plate archive database. 
         The complete database is available in electronic form upon request. } 
\label{tbl-3}
\begin{tabular}{rrrrrrrrr}
\hline
\hline
\noalign{\smallskip}
Plate nr. & Aster Nr & Aster Name & $\alpha$ & $\delta$ & Aster Type & $V$     & $\sigma$ & Dist \\
(new series) &       &            &  (J2000) &  (J2000) & (SkyBoT) & (app mag) & ($\arcsec$) & ($\arcsec$) \\
\hline
\noalign{\smallskip}
90 & - & 2002 FG7 & 15.510683424 & -6.21185735 & Amor I & 21.8 & 0.100 & 2987.286 \\
2610 &  - & 2001 YV3 & 12.50622275 & 5.23994740 & NEA & 24.6 & 105.059 & 767.524 \\
3985 &  11500 & 1989 UR & 23.75836392 & 27.79120758 & Amor I & 18.8 & 31.407 & 3154.208  \\
4592 &  89958 & 2002 LY45 & 15.48435510 & -2.69252040 & NEA & 19.7 & 22.043 & 1879.372 \\
6590 & 4660 & Nereus & 22.51019815 & -6.51922595 & Amor I & 19.6 & 0.385 & 3790.449  \\
\hline
\hline
\end{tabular}
\end{center}
\end{table*}

\begin{table*}[!t]
\begin{center}
\caption{The number of findings of asteroids in the Bucharest plate archive 1952-2005} 
\label{tbl-4}
\begin{tabular}{lrrr}
\hline
\hline
\noalign{\smallskip}
Asteroid class     & Total nr. findings & Candidate $V\le15$ findings \\
\hline
\noalign{\smallskip}
PHA prec         &     715 &      3  \\
PHA recov        &      24 &      9  \\
NEA prec         &   2,088 &      8  \\
NEA recov        &     167 &     85  \\
NUM prec         & 385,913 &    248  \\
NUM recov        &  39,977 &  7,734  \\
UNNUM (prec+rec) & 349,753 &      3  \\
\hline
\hline
\end{tabular}
\end{center}
\end{table*}

\begin{table*}[!t]
\begin{center}
\caption{NEAs and PHAs findings in the Bucharest plate archive 1952-2005 considering a limiting magnitude $V=15$. 
Two types of findings were listed in the 3rd column, "P" to denote planned observations and "S" to denote 
occasional findings. } 
\label{tbl-5}
\begin{tabular}{llrrrrr}
\hline
\hline
\noalign{\smallskip}
Asteroid Class & Asteroid Name & Type Occ & Plate Nr/   & Date & Exp Time         & Ast Mag \\
               &               &          & (nr plates) &      & (exp$\times$min) & $V$     \\
\hline
\noalign{\smallskip}
PHA prec         &  2001 WN5  & S &  4371 & Oct 1963 & 3x5 & 13.5 \\
...              &  2004 HK33 & S &  4890 & Jun 1965 &  15 & 14.6 \\
...              &  1994 CN2  & S &  4428 & Nov 1963 & 3x5 & 14.8 \\
\hline
PHA recov        & Geographos & P &  (7p) & Sep 1969 & various & 12.2 \\
...              & Toutatis   & P &  (2p) & Jan 1993 & 8x8 & 13.1 \\
\hline
NEA prec         & 1998 KU    & S &  7139 & Oct 1971 & 14,10 & 13.9 \\
...              & Eric       & S &  8386 & Oct 1973 & 2x6   & 13.9 \\
...              & Oze        & S & 10064 & Dec 1985 & 15x10 & 14.5 \\
...              & 2000 CN101 & S &  5873 & Apr 1968 & 2x7 & 14.6 \\
...              & 2000 CN101 & S &  5865 & Apr 1968 & 2x7 & 14.6 \\
...              & 1998 XB    & S &  5150 & Nov 1965 & 3x4 & 14.8 \\
...              & Jason      & S &  6502 & Oct 1970 & 3x4 & 14.8 \\
...              & Jason      & S &  6510 & Oct 1970 & 3x4 & 14.8 \\
\hline
NEA recov        & Eros       & P & (53p) &  1972-75 &  various &  7.8-13.3 \\
...              & Ganymed    & P & (19p) &  1972-73 &  various & 11.2-12.9 \\
...              & Alinda     & P &  (9p) &  1970-74 &  various & 10.2-12.5 \\
...              & Tora       & P &  7717 & Jul 1972 &  2x14 & 12.6 \\
...              & Boreas     & S &   460 & Sep 1953 & 30,10 & 14.2 \\
...              & Seleucus   & S &  9541 & Jun 1982 &  2x10 & 14.2 \\
...              & Ivar       & S &  5498 & Apr 1967 &  2x5  & 14.6 \\
\hline
\hline
\end{tabular}
\end{center}
\end{table*}



\begin{thebibliography}{}
   \bibitem[AIRA 2008]{air08} The Astronomical Institute of the Romanian Academy (AIRA): 2008, http://aira.astro.ro
   \bibitem[Berthier et al. 2006]{ber06} Berthier, J. et al.: 2006, {\it SkyBoT, a new VO service to identify Solar System objects}, 
                          Astronomical Data Analysis Software and Systems, XV ASP Conference Series, vol. 351, pag. 367
   \bibitem[Binzel, 2000]{bin00} Binzel, R. P.: 2000 Planet. Space Sci. 48, 297
   \bibitem[Boattini et al. 2001]{boa01} Boattini, A., et al.: 2001, Astronomy \& Astrophysics, 375, 293
   \bibitem[Boattini et al. 2003]{boa03} Boattini, A., et al.: 2003 Earth, Moon and Planets, 93, 239
   \bibitem[Bocsa, 2008]{boc08} Bocsa, G.: 2008, private communication
   \bibitem[Chapman, 1994]{cha94} Chapman, C. R., Morrison, D.: 1994, Nature, 367, 33
   \bibitem[Curelaru, 2007]{cur07} Curelaru, L.: 2007, private communication
   \bibitem[Gleason et al. 2004]{gle04} Gleason, A. E. et al.: 2004, {\it 2004 MN4}, Minor Planet Electr. Circ. Y70, 70
   \bibitem[Hahn, 2002]{hah02} Hahn, G.: 2002, {\it DLR-Archenhold Near Earth Objects Precovery Survey (DANEOPS)}, 
                          http://earn.dlr.de/daneops/
   \bibitem[IAU, 2008]{iau08} IAU: 2008, Minor Planet Center, 
                          http://www.cfa.harvard.edu/iau/mpc.html
   \bibitem[IMCCE, 2008]{imc08} IMCCE: 2008, Observatoire de Paris, {\it EURONEAR}, http://euronear.imcce.fr
   \bibitem[Meeus, 1999]{mee99} Meeus, J.: 1991, {\it Astronomical Algorithms}, Willmann-Bell
   \bibitem[Monet et al. 1998]{mon98} Monet, D., et al.: 1998, {\it USNO-A V2.0, A Catalog of Astrometric Standards}
   \bibitem[Sansuario and Arratia, 2008]{san08} Sansuario, M. E., Arratia, O.: 2008, Earth Moon Planet, 102, 425
   \bibitem[Steel, 1998]{ste98} Steel, D., et al.: 1998 Australian Journal of Astronomy, 7, 67
   \bibitem[Tsvetkov, 1991]{tsv91} Tsvetkov M. K., IAU Commission 9 Working Group on Wide-Field Imaging,
                        Newsletter No. 1, p. 17. http://www.skyarchive.org/wgss\_newsletter/issue1/wfpa.pdf
   \bibitem[Tsvetkov, 2005]{tsv05} Tsvetkov, M. K.: 2005, {\it Plate Content Digitisation, Archive Mining and Image Sequence Processing},   
                           Astro workshop, Sofia, 2005, ISBN-10 954-580-190-5, p. 10-41.
   \bibitem[Vaduvescu and Curelaru 2007]{vad07} Vaduvescu, O., Curelaru, L.: 2007, {\it Mining NEAs and Asteroids in the CFHTLS}, 
                          CFHT Users Meeting, 9-11 May 2007, Marseille
   \bibitem[Vaduvescu et al. 2008]{vad08} Vaduvescu, O. et al.: 2008, Planet. Space Sci. 56, 1913
   \bibitem[Vass, 1994a]{vas94a} Vass, G. et al.: 1994, Romanian Astronomical Journal, 4, 2, 179
   \bibitem[Vass, 1994b]{vas94b} Vass, G. et al.: 1994, Romanian Astronomical Journal, 4, 2, 183
\end{thebibliography}
\end{document}